\documentclass[10pt,twocolumn]{article}

\usepackage[utf8]{inputenc}
\usepackage{textgreek}
\usepackage{amsmath}
\usepackage{amssymb}
\usepackage{amsthm}
\usepackage{titlesec}
\usepackage{cite}
\usepackage{units}
\usepackage{booktabs}
\usepackage{graphicx}
\usepackage{epsfig}
\usepackage{psfrag}
\usepackage{stmaryrd}
\usepackage[colorlinks=false]{hyperref}
\usepackage[format=plain,labelfont=it]{caption}
\usepackage[left=1.5cm,right=1.5cm,top=2cm,bottom=2cm]{geometry}
\usepackage{adjustbox}
\usepackage{tikz}
\usepackage{tikzit}

\tikzstyle{Vertex}=[fill=white, draw=black, shape=circle, tikzit category=Graph]

\tikzstyle{Arrow}=[->, >=latex]

\usetikzlibrary{arrows}
\usetikzlibrary{shapes}
\usetikzlibrary{decorations.pathmorphing}
\usetikzlibrary{decorations.pathreplacing}
\usetikzlibrary{decorations.shapes}
\usetikzlibrary{decorations.text}
\usetikzlibrary{positioning, shapes}
\usetikzlibrary{calc}
\tikzset{
	block/.style = {draw, fill=white, rectangle, minimum height=3em, minimum width=4.5em},
	tmp/.style  = {coordinate}, 
	sum/.style= {draw, fill=white, circle, node distance=2cm},
	input/.style = {coordinate},
	output/.style= {coordinate}
	pinstyle/.style = {pin edge={to-,thick,black}}
}
\newlength{\dhatheight}

\usepackage[labelformat=parens,labelfont=normal]{subcaption}
\DeclareCaptionSubType*[alph]{figure}

\titleformat{\section}{\centering\normalfont\scshape}{\Roman{section}.}{5pt}{}
\titleformat{\subsection}{\normalfont\it}{\Alph{subsection}.}{5pt}{}
\titleformat{\subsubsection}{\normalfont\it}{\hspace{4mm}\arabic{subsubsection})}{5pt}{}

\newcommand\infoFootnote[1]{%
	\begingroup
	\renewcommand\thefootnote{}\footnote{#1}%
	\addtocounter{footnote}{-1}%
	\endgroup}

\newcommand{\R}{\mathbb{R}} 
\newcommand{\N}{\mathbb{N}} 

\newcommand{\Z}{\mathbb{Z}}

\newcommand{\Gc}{\mathcal{G}}
\newcommand{\Ec}{\mathcal{E}}
\newcommand{\Vc}{\mathcal{V}}
\newcommand{\Nc}{\mathcal{N}}

\newcommand{\Jc}{\mathcal{J}}

\newcommand{\mub}{\boldsymbol{\mu}}

\newcommand{\Sigmab}{\boldsymbol{\Sigma}}

\newcommand{\bb}{\boldsymbol{b}}
\newcommand{\cb}{\boldsymbol{c}}
\newcommand{\yb}{\boldsymbol{y}}

\newcommand{\eb}{\boldsymbol{e}}
\newcommand{\vb}{\boldsymbol{v}}

\newcommand{\db}{\boldsymbol{d}}

\newcommand{\xb}{\boldsymbol{x}}

\newcommand{\zb}{\boldsymbol{z}}
\newcommand{\pb}{\boldsymbol{p}}

\newcommand{\Ab}{\boldsymbol{A}}
\newcommand{\Bb}{\boldsymbol{B}}

\newcommand{\Lb}{\boldsymbol{L}}

\newcommand{\Ib}{\boldsymbol{I}}

\newcommand{\Pb}{\boldsymbol{P}}

\newcommand{\oneb}{\boldsymbol{1}}
\newcommand{\zerob}{\boldsymbol{0}}

\newcommand{\modq}{\,\,\mathrm{mod}\,\,q}

\newcommand{\kIter}{k_{\text{iter}}}
\newcommand{\kReset}{k_{\text{res}}}

\newcommand{\Enc}{\mathsf{Enc}}
\newcommand{\Dec}{\mathsf{Dec}}

\newcommand{\round}[1]{\left\lfloor#1\right\rceil}

\DeclareMathOperator*{\argmin}{arg\,min}

\title{\LARGE \bf Encrypted distributed state estimation via affine averaging
}
\author{Nils Schl\"uter$^\ast$, Philipp Binfet$^\ast$, Junsoo Kim, and Moritz Schulze Darup%
}
\date{}

\begin{document}
	\maketitle
	
	\textbf{\textit{Abstract}.} {\bf 
		Distributed state estimation arises in many applications such as position estimation in robot swarms, clock synchronization for processor networks, and data fusion. One characteristic is that agents only have access to noisy measurements of deviations between their own and neighboring states. Still, estimations of their actual state can be obtained in a fully distributed manner using algorithms such as affine averaging. However, running this algorithm, requires that the agents exchange their current state estimations, which can be a privacy issue (since they eventually reveal the actual states).  To counteract this threat, we propose an encrypted version of the affine averaging algorithm in this paper. More precisely, we use homomorphic encryption to realize an encrypted implementation, where only one ``leader'' agent has access to its state estimation in plaintext. One main challenge (which often arises for recursive encrypted computations) is to prevent overflow w.r.t.~the bounded message space of the cryptosystem. We solve this problem by periodically resetting the agents' states with the help of the leader. We study the resulting system dynamics with respect to different reset strategies and support our findings with extensive numerical simulations.
	}
	\infoFootnote{$^\ast$ N. Schl\"uter and P. Binfet equally share the first authorship. N. Schl\"uter, P. Binfet, and M. Schulze Darup are with Department of Mechanical Engineering, TU Dortmund University, Germany.
		E-mails:  \{nils.schlueter, philipp.binfet, moritz.schulzedarup\}@tu-dortmund.de.
		\\
		J. Kim is with the Department of Electrical and Information Engineering, Seoul National University of Science and Technology, Korea. E-mail: junsookim@seoultech.ac.kr}
	\infoFootnote{\hspace{-1.5mm}$^\dagger$ This paper is a \textbf{preprint} of a contribution to the 61st IEEE Conference on Decision and Control 2022.}
	
	\section{Introduction}
	With an increasing use of network resources, the impact of distributed computation schemes 
	is steadily growing. One characteristic of such schemes is that multiple agents communicate with each other and exchange data, which is private in many applications~\cite{li2009privacy}. Hence, ensuring confidentiality of the processed data is an important and common challenge.
	
	One approach to protect data during distributed computations is differential privacy~\cite{nozari2017differentially,huang2012differentially,mo2016privacy}. There, noise is injected into the private data, which results in a fundamental tradeoff between privacy and computational accuracy. Alternative approaches, which do not suffer from this tradeoff, build on homomorphic encryption (HE) or secret sharing. Both techniques allow computations over encrypted (or securely shared) data. With HE, the data is protected by the cryptosystem's security. As a result, the decryption key is necessary to obtain information from the ciphertexts, whereas in secret sharing the data is split up between computation parties. Then, the underlying assumption is that these parties are non-colluding in order to avoid an unwanted reconstruction of the secret. Approaches towards the development of distributed privacy-preserving protocols, building on these technologies, are presented in \cite{darup2020encrypted,Alexandru2019_CDC,Ruan2019_TAC,SchulzeDarup2019_LCSS,hadjicostis2020privacy,li2010secure,lee2020distributed,Kim2019encrypted,tjell2020private}.
	
	In this work, we opted for an HE-based solution, which often requires tailored reformulations of an algorithm. Such reformulations have been realized for distributed control and average consensus problems in~\cite{Alexandru2019_CDC,Ruan2019_TAC,SchulzeDarup2019_LCSS}, where the interaction between agents is secured, and each agent has the ability to decrypt aggregated data. Furthermore, fully encrypted implementations of private data aggregation can, for example, be found in~\cite{li2010secure,lee2020distributed}, and the problem of average (ratio) consensus via HE is investigated in~\cite{hadjicostis2020privacy}. A special problem arises when HE cryptosystems are used to evaluate an a-priori unknown amount of iterations. This is because without (costly) ``bootstrapping'', which is unsuitable for computationally limited agents, only a limited amount of additions and/or multiplications is offered, that has been a common concern in the field.
	
	Affine averaging~\cite{bolognani2010consensus,bullo2022lectures,barooah2007estimation}, which is applicable in various setups such as robot swarms or data fusion, is an iterative and distributed algorithm. Although agents need to exchange their states during its execution, a tailored data privacy protocol for affine averaging has (to the best of our knowledge) not yet been addressed in the literature. Therefore, we propose an encrypted realization of affine averaging with the help of HE in this paper. Of course, the problem of limited operations arises not only here. Some approaches such as~\cite{Murguia2020,SchulzeDarup2020_IFAC} build on decryption and reencryption and thereby ``resetting'' the iteration variable, which is what we adapt here. More precisely, we consider periodic resets of the agents' states with help of the leader agent, who has access to plaintext data. This way the leader can make use of its estimate while the other agents' states are protected. To realize such an update, we first find a tree subgraph by the method in~\cite{madden2002tag} of the communication graph. This defines how each agents' state is reset by means of the encrypted leader state and encrypted relative state information after a certain amount of iterations. Our novel scheme allows the encrypted evaluation of arbitrarily many iterations at the cost of quantization and reset errors, which we analyze mathematically and by a simulation study.
	
	The remaining paper is organized as follows. Section~\ref{sec:problem-statement} introduces the problem setup and the affine averaging algorithm, respectively. Then, we turn our focus towards an encrypted evaluation in Section~\ref{sec:encrytpedeval}. This is followed by an error analysis of the resulting dynamics in Section~\ref{sec:error} and an illustration of our method in Section~\ref{sec:numerical}. Finally, we end this paper with conclusions and an outlook in Section~\ref{sec:outlook}.
	
	\textit{Notation.} We denote the sets of real, integer, natural (including $0$), and positive integer numbers by $\R$, $\Z$, $\N$, and $\N_+$, respectively. The modulo operation $z \modq:=z-q\lfloor z/q \rfloor$ is frequently used, where $\lfloor\cdot \rfloor$ refers to the floor function. Further, we use $\lfloor\cdot \rceil$  for rounding to the nearest integer. Next, we write $v \sim \Nc(\mu,\sigma^2)$ for a scalar random variable that is normally distributed with mean $\mu$ and standard deviation~$\sigma$ and, analogously, $\vb \sim \Nc(\mub,\Sigmab)$ denotes a random vector with mean $\mub$ and covariance matrix $\Sigmab$. Finally, $\oneb_n$ is the vector of ones of dimension $n$, $\zerob_n$ is the corresponding zero vector, and $\Ib_n$ is the identity matrix of dimension $n\times n$.
	
	\section{Problem setup and plaintext solutions}\label{sec:problem-statement}
	We consider $n \in \N_+$ agents with unknown scalar states $x_1,\dots,x_n$ reflecting, e.g., their positions. The agents can exchange information via an (initially) undirected communication graph $\Gc$ specified by its vertices $\Vc:=\{1,\dots,n\}$ and its edges $\Ec \subset \Vc \times \Vc$. We further assume that all agents connected via an edge once measure their relative position, subject to some measurement noise. More formally, each agent~$i$ obtains the measurements
	\begin{equation}
		\label{eq:measurements}
		y_{ij}=x_i - x_j + v_{ij}
	\end{equation}
	with $v_{ij}\sim \Nc(0,\sigma_{ij}^2)$ for each neighboring agent $j \in \Jc_i$. For simplicity, we assume that the symmetries $\sigma_{ji}=\sigma_{ij}$ and $v_{ji}=-v_{ij}$ apply as in~\cite[Sect.~9.5]{bullo2022lectures} although extensions to asymmetric setups are straightforward~\cite{bolognani2010consensus}. The task now is to find an estimation $\hat{\xb} \in \R^n$ of the agents' states that best explains the obtained measurements. In order to specify the task, we first note that only one measurement per edge $\{i,j\}$ has to be considered, due to $y_{ij}=-y_{ji}$. Hence, all relevant measurements can be summarized as
	\begin{equation}
		\label{eq:yBxv}
		\yb=\Bb^\top \xb + \vb,
	\end{equation}
	where $\Bb \in \{-1,0,1\}^{n\times m}$, with $m=|\Ec|$, is an oriented incidence matrix of $\Gc$ and where $\vb \in \R^m$ reflects the noises $v_{ij}$ associated with the entries $y_{ij}$ in $\yb \in \R^m$. More compactly, we have $\vb \sim \Nc(\zerob,\Sigmab)$ with the covariance matrix  $\Sigmab=\mathrm{diag}\left(\{\sigma_{ij}^2\}_{\{i,j\}\in\Ec}\right)$. The task now becomes finding $\hat{\xb}$ such that $\yb \approx \Bb^\top \hat{\xb}$. Existing centralized and distributed solution schemes are briefly summarized in the following.
	
	\subsection{Noise-optimal centralized solution}
	An optimal interpretation of the noisy measurements can be obtained from solving
	\begin{equation}
		\label{eq:centralizedProblem}
		\hat{\xb}^{\ast}
		=\argmin_{\hat{\xb}} \left\| \Bb^{\top} \hat{\xb}-\yb\right\|_{\Sigmab^{-1}}^{2} \quad \text{s.t.} \quad \oneb_n^\top \hat{\xb} = c, 
	\end{equation}
	i.e., from minimizing the squared deviations ${\Bb^{\top} \hat{\xb}-\yb}$ weighted by the factors $1/\sigma_{ij}^2$ subject to the average constraint specified by $c\in\R$. Remarkably, the constraint in~\eqref{eq:centralizedProblem} guarantees a unique solution $\hat{\xb}^{\ast}$. Typically, one chooses $c=0$, which leads to the mean-free solution
	\begin{equation}
		\label{eq:centralizedSolution}
		\hat{\xb}^{\ast}=\Lb^{+}\Bb\Sigmab^{-1} \yb,
	\end{equation}
	where $\Lb^{+}$ denotes a pseudoinverse of the Laplacian matrix $\Lb=\Bb \Sigmab^{-1} \Bb^{\top}$ \cite[Lem.~9.8]{bullo2022lectures}. We refer to \eqref{eq:centralizedSolution} as the centralized solution since it requires the knowledge of all measurements $\yb$, the noise distribution in terms of $\Sigmab$, and the entire communication graph. 
	
	\subsection{Distributed solution via affine averaging}
	The problem becomes more challenging when a distributed solution is desired. However, a solution can still be obtained by evaluating the so-called affine averaging algorithm
	\begin{equation}
		\label{eq:affineAveraging}
		\hat{x}_i(k+1)=\hat{x}_i(k)-\alpha \sum_{j \in \Jc_i} \frac{1}{\sigma_{ij}^2}  \left(\hat{x}_i(k) - \hat{x}_j(k) - y_{ij} \right)
	\end{equation}
	on each agent, where $\alpha$ is a suitable adjustment factor (specified below). 
	Conceptually, \eqref{eq:affineAveraging} reflects iterative adjustments of the agents' states based on weighted ``edge errors''. 
	Clearly, by sorting terms, \eqref{eq:affineAveraging} can equivalently be stated as
	\begin{equation}
		\label{eq:affineAveragingDynamics}
		\hat{x}_i(k+1)= a_{ii} \hat{x}_i(k) + \sum_{j \in \Jc_i} a_{ij} \hat{x}_j(k) + b_{i} 
	\end{equation}
	with the coefficients
	\begin{equation}
		\label{eq:coefficients}
		a_{ii}=1- \sum_{j \in \Jc_i} \frac{\alpha}{\sigma_{ij}^2},\,\,\,\,  
		a_{ij}= \frac{\alpha}{\sigma_{ij}^2}, \,\,\,\,\text{and}\,\,\,\,
		b_i= \sum_{j \in \Jc_i} \frac{\alpha}{\sigma_{ij}^2}  y_{ij}.
	\end{equation}
	Finally, while affine averaging is typically applied in a distributed fashion, studying the condensed dynamics
	\begin{equation}
		\label{eq:affineAveragingCondensed}
		\hat{\xb}(k+1)=\Ab \hat{\xb}(k) + \bb 
	\end{equation}
	is useful for analysis, where $\Ab=\Ib_n - \alpha \Lb$ and $\bb=\alpha \Bb \Sigmab^{-1} \yb$. For instance, one can deduce from \eqref{eq:affineAveragingCondensed} that convergence to $\hat{\xb}^\ast$ is guaranteed if and only if (i) $\Gc$ is connected, (ii) $\oneb_n^\top \hat{\xb}(0)=0$ and (iii) $\alpha \in (0,2/\lambda_1(\Lb))$, where $\lambda_1(\Lb)$ denotes the largest eigenvalue of $\Lb$ (cf.~\cite[Lem.~9.9]{bullo2022lectures} and \cite[Sect.~4.1]{xiao2004fast}).
	
	\section{Encrypted distributed state estimation}
	\label{sec:encrytpedeval}
	\subsection{Security and privacy specifications}
	\label{subsec:securitySpecification}
	We assume in the following that the multi-agent system consists of the leader agent $i=1$ and $n-1$ ``followers'' and that only the leader should learn an estimation of its state~$\hat{x}_{1}$; ideally, without learning anything about the follower's states. Although the leader stands out in terms of the desired data access, we further assume that the computational power of all agents is equal and limited. In particular, collecting the measurements $\yb$ at the leader and computing \eqref{eq:centralizedSolution} is undesired or even intractable. As a consequence, we focus on a secure implementation of the distributed affine averaging~\eqref{eq:affineAveraging} that does not reveal $\hat{x}_i(k)$ to the followers $i \in \{2,\dots,n\}$. 
	Finally, we assume that each agent $i$ knows $\alpha$ as well as $\sigma_{ij}$ and has access to the measurements $y_{ij}$ related to its neighbors~$j\in \Jc_i$.
	
	Still, before proposing a suitable secure implementation, we briefly note that knowledge of $\hat{x}_{1}$ in combination with the measurements $y_{1j}$ for each $j \in \Jc_1$ allows to estimate the states of the leader's neighbors according to $\hat{x}_j \approx \hat{x}_1+y_{1j}$. Moreover, aiming for $\hat{\xb}^\ast$, meaningful state estimations are restricted to the constraint $\oneb_n^\top \hat{\xb}=0$. Hence, $\sum_{j=2}^n \hat{x}_j \approx -\hat{x}_1$ is unavoidably revealed to the leader as a consequence of the problem setup. Nonetheless, for most multi-agent systems this information leakage is small and, hence, acceptable from our point of view.
	
	\subsection{Encrypted affine averaging}
	\label{subsec:encryptedAffineAveraging}
	The backbone of the desired secure state estimation will be an encrypted implementation of \eqref{eq:affineAveraging} or, equivalently, of~\eqref{eq:affineAveragingDynamics}. More specifically, each agent $i$ will evaluate~\eqref{eq:affineAveragingDynamics} with encrypted states using a cryptosystem provided by the leader. Since suitable cryptosystems typically operate on integer data, we initially introduce the integer-based representation
	\begin{equation}
		\label{eq:integerDynamics}
		z_i(k+1)= \round{s a_{ii}} z_i(k) + \sum_{j \in \Jc_i} \round{s a_{ij}} z_j(k) + \round{s^k} \round{s^2 b_i}
	\end{equation}
	of~\eqref{eq:affineAveragingDynamics}, where $z_i(0)= \round{s \hat{x}_i(0)}\in\Z$ and where $s \in [1, \infty)$ refers to a scaling factor. Clearly, for a sufficiently large choice of~$s$, the relation
	\begin{equation}
		\label{eq:xiRecovery}
		\hat{x}_i(k) \approx \frac{1}{s^{k+1}}  z_i(k)
	\end{equation}
	applies which allows recovering the agents' states. In the following, we specify the encrypted implementation of~\eqref{eq:integerDynamics} using additively HE. We note, however, that various cryptosystems which support secure computations could be applied. Now, an additively HE scheme basically provides the following three features.
	\begin{enumerate}
		\item An encryption $\Enc(\cdot)$, which allows encrypting data from $\Z_q:=\{0,\dots,q-1\}$ with $q\in \N_+$ and a decryption $\Dec(\cdot)$, which decrypts the corresponding ciphertexts according to $\Dec(\Enc(z^\prime))=z^\prime$ for every $z^\prime \in \Z_q$.
		\item An operation ``$\oplus$'' enabling encrypted~additions according to $\Dec\left(\Enc(z^\prime_
		{\mathrm{I}}) \oplus \Enc(z^\prime_\mathrm{II})\right)= z^\prime_\mathrm{I} + z^\prime_\mathrm{II} \modq$.
		\item An operation ``$\otimes$'' enabling partially encrypted multiplications via ${\Dec\left( z^\prime_\mathrm{I} \otimes \Enc(z^\prime_{\mathrm{II}})\right)= z^\prime_\mathrm{I} z^\prime_\mathrm{II} \modq}$.
	\end{enumerate}
	With these three features at hand, an encrypted implementation of~\eqref{eq:integerDynamics} can be realized as follows. First, each agent $i$ is aware of its measurements $y_{ij}$ and the corresponding accuracies in terms of $\sigma_{ij}$. Hence, it can compute the coefficients $\round{s a_{ij}}$ for all its neighbors and $\round{s^{2} b_i}$. Second, as required by the HE scheme, only messages in $\Z_q$ can be processed. A mapping to $\Z_q$ is realized by applying modulo~$q$ (the reconstruction from $\Z_q$ is discussed below). For the states, we then obtain $z_j^\prime(k):=z_j(k)\modq$. Next, a secure computation of the products $\round{s a_{ij}}\modq\otimes \Enc\left(z_j^\prime(k)\right)$ (including $i=j$) can be performed. Finally, all resulting terms and $\Enc\left(\round{s^k} \round{s^{2} b_i}\modq\right)$ can be added up using the operation~``$\oplus$''. Because HE schemes are typically public-key cryptosystems, the encryption procedure is publicly available and only the key for decryption is a secret; in our setup, it is only known to the leader.
	
	While the encrypted implementation is indeed simple, one has to take into account that the scaling factor associated with $z_i(k)$ is increasing with each iteration of \eqref{eq:integerDynamics} as also apparent from~\eqref{eq:xiRecovery}. This becomes an issue if  $z_1(k)< -q/2$ or $q/2 \leq z_1(k)$, because it implies an overflow of $\Z_q$ and thus an erroneous reconstruction of the leader's state. Conversely, a sufficient condition\footnote{The condition is also necessary for odd $q$, and it only neglects the feasible state $z_1(k)=-q/2$ for even $q$.} to avoid overflow is $|z_1(k)|< q/2$. Then, in order to recover the state, one can use the reconstruction
	\begin{equation}
		\label{eq:modinv}
		z_1(k)=\left\{ \begin{array}{ll}
			z_1^\prime(k)   & \text{if}\,\,z_1^\prime(k)< q/2, \\
			z_1^\prime(k)-q   &  \text{otherwise.}
		\end{array}\right.
	\end{equation}
	Now, let us assume that a certain $s$ is required for accuracy reasons, that $q$ has been chosen with respect to the security demands, and that (possibly conservative) bounds on $\hat{x}_1(k)$ are known in terms of $|\hat{x}_1(k)| < \bar{x}_1$ for every $k\in\N$. Then, \eqref{eq:xiRecovery} suggests that the maximal number of iterations $\kIter\in \N$, that can be carried out without overflow, is limited by the condition
	\begin{equation}
		\label{eq:kIterCondition}
		s^{\kIter+1} \left(\bar{x}_1+\delta(\kIter)\right) < q/2,
	\end{equation}
	where $\delta:\N \rightarrow \R$ reflects approximation errors as specified in Section~\ref{subsec:errrosIntegerBased}. Since it will turn out that $\delta$ is non-decreasing, guaranteeing \eqref{eq:kIterCondition} prevents an overflow for every $k \in \{1,\dots,\kIter\}$ (where we exclude $k=0$, since $z_1(0)=\round{s\hat{x}_1(0)}$ is known to the leader anyway). In other words, the leader can reliably reconstruct $z_1(k+1)$ from ${z_1^\prime(k+1)}$ for every $k \in \{0,\dots,\kIter-1\}$, where 
	\begin{align*}
		\nonumber
		z_1^\prime(k+1)&=\round{s a_{11}} z_1(k)+\sum_{j\in\Jc_1} \round{s a_{1j}} \Dec\left(\Enc(z_j^\prime(k))\right) \\
		&\quad + \round{s^k} \round{s^2 b_{1}} \modq. 
	\end{align*}
	Remarkably, whether an overflow occurs when recovering the states $z_j(k)$ from $z_j^\prime(k)$ for $j\in \Jc_1$ (or even for the remaining followers) is irrelevant for a correct recovery of $z_1(k+1)$, which is a property of the modulo operation.
	
	\subsection{Enabling additional iterations through tailored resets}\label{subsec:reset-protocol}
	The accumulation of scaling factors during iterative computations is a well-known issue in encrypted control \cite{Kim2016,Cheon2018need,Murguia2020,SchulzeDarup2020_IFAC,schlueter2021stability}. A simple way to break the accumulation is a periodical reset of the integer states. 
	More precisely, the idea is to replace $z_i(\kIter)$, which approximately corresponds to $s^{\kIter+1} \hat{x}_i(\kIter)$ according to~\eqref{eq:xiRecovery}, with $z_i(\kIter) \leftarrow \round{s \hat{x}_{i,\text{res}}}$, where $\hat{x}_{i,\text{res}}$ should ideally be equal to $\hat{x}_i(\kIter)$ but may also take different values. In fact, the primary goal is to reset the scaling factor and keeping a close relation to $\hat{x}_i(\kIter)$ is only secondary.
	This prioritization explains why, for other problem setups, even ``radical'' resets with $\hat{x}_{i,\text{res}}=0$ have been proposed \cite{Murguia2020}. In fact, such a reset is appealing from a cryptographic point of view since $\Enc(z_i^\prime(\kIter)) \leftarrow \Enc(0)$ can be carried out locally without the need for interaction or decryption. In contrast, the more desirable choice $\hat{x}_{i,\text{res}}=\hat{x}_i(\kIter)$ either requires bootstrapping as in \cite{Kim2016} or ``external'' decryption and re-encryption as, e.g., in \cite{SchulzeDarup2020_IFAC}.
	
	For our setup, the radical resets $\hat{x}_{i,\text{res}}=0$ do not make sense because this will discard all previous iteration results. Then, each subsequent computation phase will (again) be initialized with $\hat{x}_{i,\text{res}}=0$, which results in a limit cycle. Next, taking into account that the encrypted implementation of~\eqref{eq:integerDynamics} can be realized with relatively lightweight cryptosystems (such as additively HE) and that the computational power of the agents is limited, applying a demanding HE scheme which offers bootstrapping is also not an option. Thus, we will reset the followers using ``external'' decryption carried out by the leader. However, we will not simply forward all $\Enc\left(z_i^\prime(\kIter)\right)$ to the leader, since this would clearly violate the privacy. Instead, we will realize resets based on accumulated measurements $y_{ij}$, which do not reveal more than the already discussed insights \mbox{(see Sect.~\ref{subsec:securitySpecification})}. 
	
	In order to specify a tailored reset procedure, we initially note that the noise-optimal solution \eqref{eq:centralizedSolution} allows specifying the ``fictitious'' measurements $\hat{\yb}^\ast:=\Bb^\top\hat{\xb}^\ast$, which perfectly and consistently reflect the various differences $\hat{x}_i^\ast-\hat{x}_j^\ast$. Hence, we can describe each of the followers' states $\hat{x}_i^\ast$ by starting at $\hat{x}_1^\ast$ and subtracting edge measurements along a finite and directed path from node $1$ to $i$. More precisely, we find
	\begin{equation}
		\label{eq:xlOpt_via_x1Opt_and_yijOpt}
		\hat{x}_i^\ast = \hat{x}_1^\ast - \hat{y}_{1j}^\ast - \dots  -\hat{y}_{li}^\ast,   
	\end{equation}
	where $j$ and $l$ reflect the second and the last node on the path, respectively. Here, $\hat{d}^\ast_i:=\hat{y}_{1j}^\ast + \dots +\hat{y}_{jl}^\ast$ can be interpreted as the signed distance between $\hat{x}_1^\ast$ and $\hat{x}_i^\ast$ since we obviously have
	$\hat{d}^\ast_i = \hat{x}_1^\ast-\hat{x}_i^\ast$ (where a negative sign indicates $\hat{x}_1^\ast<\hat{x}_i^\ast$).
	A more technical description of the corresponding path is as follows. Let $\pb_i \in \{-1,0,1\}^m$ be such that $\Bb \pb_i = \eb_1 - \eb_i$, where $\eb_i$ denotes the $i$-th canonical unit vector in $\R^n$. Then, the $j$-th entry in $\pb_i$ specifies whether the edge corresponding to the $j$-th column of $\Bb$ is traversed in positive ($1$) or negative ($-1$) direction on the path from node $1$ to $i$ or not passed at all~($0$). Remarkably, using $\pb_i$, we obtain the compact relation $\hat{d}^\ast_i=\pb_i^\top \hat{\yb}^\ast$. 
	
	Now, assuming that the path vector $\pb_i$ has been identified for each follower  $i\in\{2,\dots,n\}$, a reset procedure can be derived as follows. First, we note that $d_i=\pb_i^\top \yb$ provides an estimation of $\hat{d}^\ast_i$ based on the actual measurements. Hence, we could also obtain estimations of $\hat{x}_i^\ast$ by evaluating ${\hat{x}_1(\kIter)+d_i}$. However, using these relations to reset the agents' states would most likely violate the zero mean restriction. More formally,
	\begin{equation}
		\label{eq:naiveReset}
		\hat{x}_1(\kIter) + \sum_{i=2}^n \left(\hat{x}_1(\kIter) - d_i \right) = n \hat{x}_1(\kIter)  - \sum_{i=2}^n d_i
	\end{equation}
	is, in general, not equal to $0$. Hence, we need to modify the procedure. A suitable modification is to replace $\hat{x}_1(\kIter)$ with
	\begin{equation}
		\label{eq:hardReset}
		\hat{x}_1(\kIter) \leftarrow \frac{1}{n}\sum_{i=2}^n d_i
	\end{equation}
	before performing the reset. Indeed, substituting~\eqref{eq:hardReset} in~\eqref{eq:naiveReset} leads to $0$ by construction. However, it is important to note that~\eqref{eq:hardReset} is not the only admissible choice. In fact, additional choices arise from distinguishing between the modification $\Delta \hat{x}_1$ to the actual leader state and the modification $\Delta \hat{x}_{\Gc}$ to the leader's state ``copy'' used to reset the followers. More formally, choosing $\Delta \hat{x}_1$ and $\Delta \hat{x}_{\Gc}$ such that
	$$
	\hat{x}_1(\kIter)-\Delta \hat{x}_1 + \sum_{i=2}^n \left(\hat{x}_1(\kIter)-\Delta \hat{x}_\Gc - d_i \right) = 0
	$$
	allows to perform the admissible resets 
	\begin{subequations}
		\label{eq:modifiedResets}
		\begin{align}
			\!\!\!\hat{x}_1(\kIter+\kReset) &\leftarrow \hat{x}_1(\kIter)-\Delta \hat{x}_1 & &\!\!\!\!\!\text{and} \\
			\!\!\!\hat{x}_i(\kIter+\kReset) &\leftarrow \hat{x}_1(\kIter)-\Delta \hat{x}_{\Gc}-d_i & &\!\!\!\!\!\forall i \in \!\{2,\dots,n\},\!\!
		\end{align}
	\end{subequations}
	where $\kReset \in \N_+$ reflects the fact that implementing the reset will require a certain number of (communication) steps (as specified below). Naturally, the squared modifications $\Delta \hat{x}_1^2$ and $\Delta \hat{x}_{\Gc}^2$ should be as small as possible. In order to derive specific choices, we combine both objectives in a cost function $\Delta \hat{x}_1^2+w\Delta \hat{x}_{\Gc}^2$ with the weighting factor $w$. For any non-negative choice of $w \in \R$ we then obtain
	\begin{equation}
		\label{eq:dxForResets}
		\begin{pmatrix}
			\Delta \hat{x}_{1} \\
			\Delta \hat{x}_{\Gc}
		\end{pmatrix} = \frac{1}{(n-1)^2+w} \begin{pmatrix}
			w \\
			n-1
		\end{pmatrix} \left(n \hat{x}_1(k_{\mathrm{iter}}) - \sum_{i=2}^n d_i \right).
	\end{equation}
	Now, some choices of $w$ refer to special cases. For instance, $w=0$ implies $\Delta \hat{x}_{1}=0$. Hence, only the copy of the leader's state is modified. Another special case is $w=n-1$, for which we find 
	\begin{equation}
		\label{eq:hardResetDeltas}
		\Delta \hat{x}_{1} =  \Delta \hat{x}_{\Gc} = \hat{x}_1(k_\mathrm{iter}) - \frac{1}{n} \sum_{i=2}^n d_i.
	\end{equation}
	Remarkably, this case is identical to the reset in \eqref{eq:hardReset}. This becomes clear from substituting \eqref{eq:hardResetDeltas} in \eqref{eq:modifiedResets} and noting that the leader's state used for both resets is identical to \eqref{eq:hardReset}. Since this reset is the only reset, where the previous iteration result $\hat{x}_1(k_{\mathrm{iter}})$ is not used, we refer to it as the \textit{hard reset}. All other resets with $w \in [0,\infty) \setminus \{n-1\}$ are called \textit{soft resets}.
	
	\subsection{Encrypted periodical resets via tree}
	In order to implement the proposed resets in an encrypted fashion, the leader requires $\sum_{i=2}^n d_i$. We note, in this context, that the paths corresponding to $\pb_2$ to $\pb_n$ establish a directed subgraph of $\Gc$, as also apparent from Figure~\ref{subfig:ex-graph-directed-sub}. Further, this subgraph can be interpreted as a tree with the leader as the root node and the followers as inner and leaf nodes (see Fig.~\ref{subfig:ex-graph-directed-sub-tree}). Now, in order to provide $\sum_{i=2}^n d_i$ to the leader, a leaf agent $i$ will send its negated measurement $y_{ij}$ to its parent node $j$, where the negation is required to compensate for traversing the directed edge in negative direction. For instance, agent $5$ in Figure~\ref{subfig:ex-graph-directed-sub-tree} will report $-y_{53}=y_{35}$ to agent $3$. The parent node will collect and add up all data from its children, add the measurement w.r.t. to its parent node and forward the accumulated data to the parent. At some point, the leader receives the accumulated data from its children, and it is easy to see that adding up the summands leads to $\sum_{i=2}^n d_i$ as desired. Thus, including its state $\hat{x}_1(\kIter)$ and the choice for $w$, the leader can prepare the resets by evaluating~\eqref{eq:dxForResets}. To implement the resets, the tree is now traversed top down. More precisely, the leader forwards $\hat{x}_1(\kIter)+\Delta \hat{x}_{\Gc}-y_{1i}$ to each of its children~$i$. The children reset their states accordingly and forward $(\hat{x}_1(\kIter)+\Delta \hat{x}_{\Gc}-y_{1i})-y_{ij}$ to their children $j$ until the leaf nodes are reached, and the reset is complete.
	
	The encrypted realization of the resets works analogously. In fact, it only differs in that the communicated data is encrypted. More precisely, during the data collection, a leaf agent $i$ will not forward $-y_{ij}$ to its parent but $\Enc(\round{-s y_{ij}} \modq)$. The parents likewise add their encrypted contributions using the operation ``$\oplus$''. 
	Finally, from the received data, the leader obtains
	\begin{equation}
		\label{eq:dSumInteger}
		\frac{1}{s} \sum_{i=2} \pb_i^\top \round{s \yb} \approx \sum_{i=2}^n d_i
	\end{equation}
	after decryption, reconstruction analogously to~\eqref{eq:modinv}, and division by $s$. Using this approximation, the leader will compute the approximations $\Delta \check{x}_{1}$ and $\Delta \check{x}_{\Gc}$ analogously to~\eqref{eq:dxForResets} with $\check{x}_{1}(\kIter):=z_1(\kIter)/s^{\kIter+1}$. Afterwards, the leader will transmit $\Enc(\round{s(\check{x}_1(\kIter)+\Delta \check{x}_{\Gc})}-\round{s y_{1i})} \modq)$ to each of its children~$i$, who add $\Enc(\round{-sy_{ij}} \modq)$ for their children $j$.
	Finally, every follower $i$ holds $$
	\Enc(\round{s(\check{x}_1(\kIter)+\Delta \check{x}_{\Gc})}-\pb_i^\top \round{s \yb} \modq)
	$$
	for updating its integer state without being able to decrypt~it.\!
	
	It remains to comment on the number of steps required to implement the resets. As apparent from the described data collection and distribution phase, the longest path from the leader to a follower determines the number of required communication steps. Clearly, the length of the longest path is equal to the height $h$ of the tree (see also Fig.~\ref{subfig:ex-graph-directed-sub-tree}). Hence, assuming that the number of evaluated iterations satisfies $\kIter\geq h$, the data collection can be carried out simultaneously to the first computation phase. It is further interesting to note that the data collection is only required once, since the measurements $\yb$ are static in our setup. Now, distributing the reset information to the outermost agent requires exactly $\kReset=h$ steps, where $\kReset$ has already been used in~\eqref{eq:modifiedResets}. Remarkably, no iterations~\eqref{eq:integerDynamics} are carried out during the resets.
	
	In summary, our encrypted affine averaging algorithm first considers $\Enc\left(\zb^\prime(0)=\zerob\right)$ and evaluates $\kIter$ of the encrypted version of \eqref{eq:integerDynamics} while simultaneously collecting~\eqref{eq:dSumInteger} at the leader. Then, depending on $z_1(\kIter)$ and the choice of $w$, it uses a reset, which requires $\kReset$ steps and leads to $\Enc\left(\zb^\prime(\kIter+\kReset)\right)$. With the ``refreshed'' states, another computation round of $\kIter$ steps follows and so forth. The algorithm terminates when no further improvements from $z_1(k)$ to $z_1(k+1)$ are observed (as specified below).
	
	\section{Error analysis}
	\label{sec:error}
	Our encrypted affine averaging algorithm clearly differs from its plaintext archetype. More specifically, two mechanisms will cause deviations. First, the integer-based representation is subject to quantization errors. Second, the periodical resets affect the convergence of the algorithm. We will briefly study both effects in the following.
	
	\subsection{Errors resulting from integer-based realization}
	\label{subsec:errrosIntegerBased}
	The behavior of the original affine averaging algorithm is typically studied based on~\eqref{eq:affineAveragingCondensed}. Also, the integer-based algorithm~\eqref{eq:integerDynamics} can be condensed to the analog form
	\begin{equation}
		\label{eq:integerDynamicsCondensed}
		\zb(k+1)=  \round{s \Ab} \zb(k) + \round{s^k} \round{s^{2} \bb}.
	\end{equation}
	Now, studying the approximation errors underlying~\eqref{eq:xiRecovery} basically aims for the identification of a function $\delta:\N \rightarrow \R$, which bounds the deviations from above according to
	\begin{equation}
		\label{eq:deltaDeviations}
		\|\zb(k)/s^{k+1} -\hat{\xb}(k)  \|_\infty \leq \delta(k).
	\end{equation}
	Clearly, \eqref{eq:affineAveragingCondensed} leads to the explicit expression
	\begin{equation}
		\label{eq:affineAveragingExplicit}
		\hat{\xb}(k)=\Ab^k \hat{\xb}(0) + \left( \sum_{j=0}^{k-1} \Ab^j \right) \bb.
	\end{equation}
	Analogously, \eqref{eq:integerDynamicsCondensed} implies 
	\begin{equation}
		\label{eq:integerDynamicsExplicit}
		\zb(k)=  \round{s \Ab}^k \zb(0) +  \left(\sum_{j=0}^{k-1} \round{s \Ab}^{j} \round{s^{k-1-j}}\right) \round{s^2 \bb}.
	\end{equation}
	In principle, we can now substitute~\eqref{eq:affineAveragingExplicit} and~\eqref{eq:integerDynamicsExplicit} in~\eqref{eq:deltaDeviations} and derive suitable overestimations. However, before analyzing the resulting expressions, we consider two scalar expressions and note that
	\begin{align}
		\label{eq:scalerDeviationHigherScaling}
		\left|\frac{\round{s^i x}}{s^i} -x\right| &\leq \frac{1}{2 s^i} \qquad \text{and} \\ 
		\label{eq:scalerDeviationHigherPower}
		\left|\frac{\round{s x}^j}{s^j} - x^j \right| & =  \left| \sum_{l=0}^{j-1} \binom{j}{l}  \Delta x^{j-l} x^{l} \right| \leq  \sum_{l=0}^{j-1} \binom{j}{l} \left|\Delta x^{j-l} \right| \left|x^l \right|\\
		\nonumber
		&\leq \sum_{l=0}^{j-1} \binom{j}{l}\left(\frac{1}{2s}\right)^{j-l} \left|x \right|^{l} = \left(\left|x \right|+\frac{1}{2 s}\right)^{j}\!\! -\left|x \right|^{j}
	\end{align}
	apply for any $x\in \R$ and $i,j \in \N$, where ${\Delta x := \round{s x}/s-x}$ and where $\binom{j}{l}$ refers to a binomial coefficient. In fact, the first relation is easy to verify and implies ${|\Delta x|\leq 1/(2s)}$, which allows deriving the second relation. As discussed next, the relations \eqref{eq:scalerDeviationHigherScaling} and~\eqref{eq:scalerDeviationHigherPower} can be extended to vectors and matrices. For instance, we obviously have $\| \round{s^2 \bb}/s^2  - \bb \|_\infty \leq 1/(2s^2)$. However, the extension of \eqref{eq:scalerDeviationHigherScaling} to matrices requires small modifications. In fact, we find $\| \round{s \Ab}/s  - \Ab \|_\infty \leq \nu/(2s)$, where $\nu$ reflects the maximal number of non-zero entries in any row of $\Ab$. For our setup, $\nu:=1+\max_{i \in \Vc} |\Jc_i|$ is typically significantly smaller than $n$ (and equals the maximal out-degree plus~$1$). Now, taking into account that the matrix $\infty$-norm is sub-multiplicative, the extension of \eqref{eq:scalerDeviationHigherPower} initially yields
	$$
	\left\|\frac{\round{s \Ab}^j}{s^j} - \Ab^j \right\|_\infty \leq \left(\left\| \Ab \right\|_\infty+\frac{\nu}{2 s}\right)^{j}\!\! -\left\| \Ab \right\|_\infty^j.
	$$
	Here, another simplification is possible because $\Ab$ is known to be double-stochastic whenever $\alpha \in (0,1/\lambda_1(\Lb)]$, which implies $\left\| \Ab \right\|_\infty=1$. We are now ready to investigate~\eqref{eq:deltaDeviations} term by term.
	Most importantly, by matching summands from~\eqref{eq:affineAveragingExplicit} and~\eqref{eq:integerDynamicsExplicit}, we find terms of the form
	\begin{equation*}
		\Delta \cb_j:=\round{s \Ab}^{j} \round{s^{k-1-j}} \round{s^2 \bb}/s^{k+1} - \Ab^j \bb.
	\end{equation*}
	Overestimating the $\infty$-norm of this term can be carried out based on the relations above. 
	To simplify the analysis, we assume $s\in\N_+$, which implies $\round{s^{k-1-j}}/s^{k+1}=1/s^{j+2}$. Using the shorthand notation $\Delta \Ab_j:={\round{s \Ab}^{j}/s^j-\Ab^j}$ and $\Delta \bb:=\round{s^2 \bb}/s^2-\bb$, we then find that $\|\Delta \cb_j\|_\infty$ is 
	\begin{align*}
		&\left\| \round{s \Ab}^{j} \round{s^2 \bb}/s^{j+2} - \Ab^j \bb \right\|_\infty \\
		=&\left\| (\Ab^j+\Delta \Ab_j)(\bb+\Delta \bb) -\Ab^j \bb \right\|_\infty\\ 
		= &\left\| \Ab^j \Delta \bb + \Delta \Ab_j \bb +  \Delta \Ab_j \Delta \bb \right\|_\infty \\
		\leq &\|  \Ab \|_\infty^j \| \Delta \bb \|_\infty + \|\Delta \Ab_j\|_\infty \left( \|\bb\|_\infty +  \| \Delta \bb \|_\infty \right) \\
		\leq &\frac{\|\Ab\|_\infty^j}{2s^2} \!+ \left(\left(\|\Ab\|_\infty\!+\frac{\nu}{2 s}\right)^{j}\!\! -\|\Ab\|_\infty^j\right)  \left( \|\bb\|_\infty \!+  \frac{1}{2s^2} \right)\\
		=&\left(\|\Ab\|_\infty+\frac{\nu}{2 s}\right)^{j} \left( \|\bb\|_\infty +  \frac{1}{2s^2} \right) - \|\Ab\|_\infty^j \|\bb\|_\infty.
	\end{align*}
	Now, taking into account that $\zb(0)=\hat{\xb}(0)=\zerob$, a valid specification for $\delta(k)$ in~\eqref{eq:deltaDeviations} and $k\in \{0,\dots,\kIter\}$ is
	\begin{equation}
		\label{eq:delta}
		\delta(k)=\sum_{j=0}^{k-1} \|\Delta \cb_j\|_\infty.
	\end{equation}
	Clearly, after the first reset, another term needs to be added to compensate for deviations between the reset integer state $\zb(\kIter+\kReset)$ and its unquantized analogue. Structurally, the corresponding terms $\round{s \Ab}^{k-\kIter-\kReset} \zb(\kIter+\kReset)$ can, however, be analyzed analogously to $\round{s \Ab}^{j} \round{s^2 \bb}$. Thus, we omit a detailed analysis due to space restrictions. We conclude this section by discussing two features of $\delta$ as in~\eqref{eq:delta}. First, $\delta$ is indeed non-decreasing, as already claimed in Section~\ref{subsec:encryptedAffineAveraging}. Second, $\lim\limits_{s\to \infty} \delta(k)=0$. Hence, by choosing $s$ large enough, we can make the deviations in~\eqref{eq:deltaDeviations} arbitrarily small.
	
	\subsection{Errors related to resets}
	The analysis of the errors related to resets requires a different methodology compared to Section~\ref{subsec:errrosIntegerBased}. In fact, the resets build on the estimated $d_i$, which are affected by the noise of the measurements. Hence, we will investigate stochastic deviations in the following. As a preparation, we investigate the relation of the actual states $\xb$ and the noise-optimal and mean-free estimations $\hat{\xb}^\ast$. To this end, we substitute \eqref{eq:yBxv} in \eqref{eq:centralizedSolution} and find
	\begin{align}
		\nonumber
		\hat{\xb}^{\ast}=\Lb^+\Bb\Sigmab^{-1} \yb &=\Lb^+\Bb\Sigmab^{-1} (\Bb^\top \xb + \vb) \\
		\label{eq:xOptNoise}
		&= \Lb^+ \Lb \xb + \Lb^+\Bb\Sigmab^{-1} \vb.
	\end{align}
	Taking into account that $\Lb^+ \Lb = \Ib_n - \frac{1}{n} \oneb_n \oneb_n^\top$  \cite[p.~104]{bullo2022lectures}, it becomes clear that the expression $\tilde{\xb}=\Lb^+ \Lb \xb$ refers to mean-free states resulting from a shift of $\xb$. Hence, $\hat{\xb}^{\ast}$ can be interpreted as a normally distributed random variable with the expected value $\tilde{\xb}$ and the covariance matrix
	\begin{align*}
		\Lb^+\Bb\Sigmab^{-1} \Sigmab (\Lb^+\Bb\Sigmab^{-1})^\top  &= \Lb^+\Bb\Sigmab^{-1} \Sigmab \Sigmab^{-1} \Bb^\top \Lb^+\\
		&= \Lb^+\Lb \Lb^+ = \Lb^+.   
	\end{align*}
	We next intend to carry out a similar analysis for the states resulting from the resets in Section~\ref{subsec:reset-protocol}. Since the hard reset (resulting for $w=n-1$) ignores the current leader state, we begin our analysis with this case. To enable a compact presentation, we introduce the vector $\db \in \R^n$ reflecting the (signed) distances $d_i$, where we set $d_1=0$, and the matrix 
	\begin{equation}
		\label{eq:pathMatrix}
		\Pb=\begin{pmatrix}
			\zerob_m & \pb_2 & \dots & \pb_n
		\end{pmatrix} \in \{-1,0,1\}^{n \times m}.
	\end{equation}
	We then find $\db =  \Pb^\top \yb$ in agreement with Section~\ref{subsec:reset-protocol}. 
	Further, the hard reset can be described as
	$$
	\hat{\xb}_{\text{hard}}=\frac{1}{n} \oneb_n \oneb_n^\top \db - \db = - \Lb^+ \Lb \db
	$$
	according to \eqref{eq:hardReset}. Now, considering that ${\Bb \pb_i=\eb_1-\eb_i}$ holds by construction, we obtain
	$$
	\Bb \Pb = \begin{pmatrix}
		\zerob_m & \eb_1-\eb_2 & \dots & \eb_1-\eb_n
	\end{pmatrix} 
	$$
	and consequently 
	\begin{equation}
		\label{eq:dNoise}
		\db = \Pb^\top \yb = (\Bb \Pb)^\top \xb + \Pb^\top\vb=\oneb_n x_1 - \xb + \Pb^\top \vb.   
	\end{equation}
	In summary, we find
	$$
	\hat{\xb}_{\text{hard}}  = - \Lb^+ \Lb \db  = - \Lb^+ \Lb (\oneb_n x_1 - \xb + \Pb^\top \vb) = \tilde{\xb} - \Lb^+ \Lb \Pb^\top \vb.
	$$
	Thus, the hard reset can likewise be interpreted as a normally distributed random variable with the expected value $\tilde{\xb}$ but with the covariance matrix
	$$
	\Lb^+ \Lb \Pb^\top \Sigmab ( \Lb^+ \Lb \Pb^\top )^\top =  \Lb^+ \Lb \Pb^\top  \Sigmab \Pb \Lb\Lb^+.
	$$
	In principle, the soft resets can be analyzed analogously. However, for brevity, we only consider the special case $w=0$. Given the leader state $\hat{x}_1$, this reset leads to
	\begin{align*}
		\hat{\xb}_{\text{soft},0}&=\begin{pmatrix}
			\hat{x}_1 \\
			\oneb_{n-1} (\hat{x}_1-\frac{1}{n-1} (n \hat{x}_1 - \oneb_n^\top \db))
		\end{pmatrix} - \db \\
		&=\begin{pmatrix}
			1 \\
			-\frac{\oneb_{n-1}}{n-1}
		\end{pmatrix} \hat{x}_1 + \left( \begin{pmatrix}
			\zerob_n^\top \\
			\frac{\oneb_{n-1}  \oneb_n^\top}{n-1}
		\end{pmatrix}   - \Ib_n \right) \db.
	\end{align*}
	Clearly, the outcome of this reset crucially depends on the state $\hat{x}_1$. In order to investigate the best possible outcome, we consider $\hat{x}_1=\eb_1^\top \hat{\xb}^\ast$ for the purpose of analysis. Now, substituting \eqref{eq:xOptNoise} and \eqref{eq:dNoise} would allow studying the mean and covariance of the soft reset for $w=0$. Since the analysis of the covariance is cumbersome, we focus on the expected value and find 
	$$
	E(\hat{\xb}_{\text{soft},0})=\begin{pmatrix}
		1 \\
		-\frac{\oneb_{n-1}}{n-1}
	\end{pmatrix} \tilde{x}_1 + \left( \begin{pmatrix}
		\zerob_n^\top \\
		\frac{\oneb_{n-1}  \oneb_n^\top}{n-1}
	\end{pmatrix}   - \Ib_n \right) (\oneb_n \tilde{x}_1 - \tilde{\xb})
	$$
	with $\tilde{x}_1$ being the first entry of $\tilde{\xb}$ and by noting that ${\oneb_n \tilde{x}_1 - \tilde{\xb}}=\oneb_n x_1 - \xb$. Taking into account that $\oneb_n^\top \oneb_n=n$, we easily derive $E(\hat{\xb}_{\text{soft},0})=\tilde{\xb}$. Hence, also the soft reset (for $w=0$) is meaningful supposed that the estimation of the leader's state is reasonable.
	\begin{figure*}[htp!]
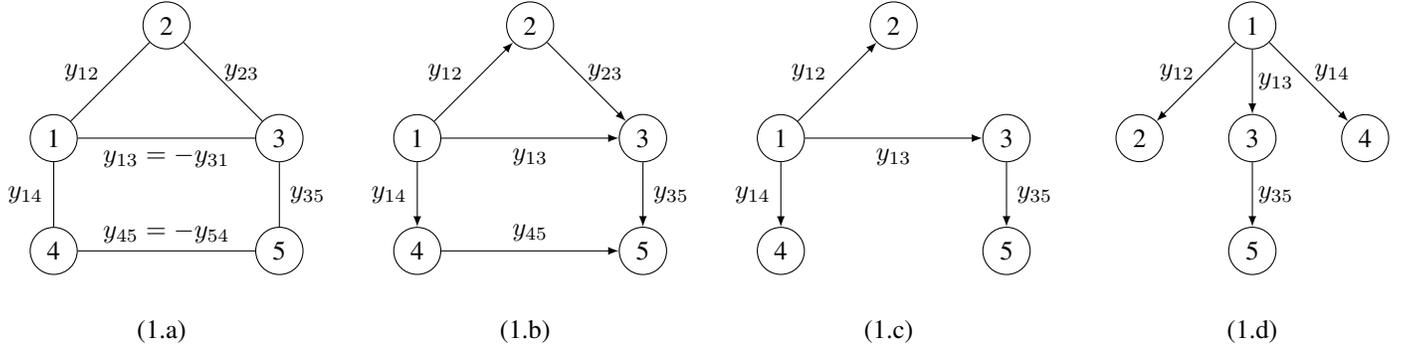

		\centering
		\subcaptionbox{\label{subfig:ex-graph-undirected}}[0.22\textwidth]{%
			\ctikzfig{Figures/example_graph_undirected}
		}
		\hfill
		\subcaptionbox{\label{subfig:ex-graph-directed}}[0.22\textwidth]{%
			\ctikzfig{Figures/example_graph_directed}
		}
		\hfill
		\subcaptionbox{\label{subfig:ex-graph-directed-sub}}[0.22\textwidth]{%
			\ctikzfig{Figures/example_graph_directed_sub}
		}
		\hfill
		\subcaptionbox{\label{subfig:ex-graph-directed-sub-tree}}[0.22\textwidth]{%
			\ctikzfig{Figures/example_graph_directed_sub_tree}
		}
		\caption{The different graph variants involved in the small-scale example. (\subref{subfig:ex-graph-undirected}) shows the initial communication graph. Note that two measurements, $y_{ij}$ and $y_{ji}$, are associated with each edge ${i,j}$ and that $y_{ij} = -y_{ji}$ holds for all edges. In (\subref{subfig:ex-graph-directed}), particular edge directions (and measurements) have been chosen. (\subref{subfig:ex-graph-directed-sub}) depicts the directed subgraph used for the reset procedure. (\subref{subfig:ex-graph-directed-sub-tree}) is the same as (\subref{subfig:ex-graph-directed-sub}) but reorganized to a tree-like structure with the leader as its root.}
		\label{fig:example-small-graphs}
	\end{figure*}
	
	\section{Numerical benchmark}
	\label{sec:numerical}
	We illustrate our results with an academic small-scale example with $n=5$ agents and randomly generated medium-scale problems with $n \in \{10,11,\dots,100\}$.
	
	\subsection{Small-scale example}
	In order to clarify the notation and the crucial steps of the proposed scheme, we consider the undirected graph in  Figure~\ref{subfig:ex-graph-undirected} with $n=5$ nodes and $m=6$ edges. Each node represents an agent~$i$ who holds noisy measurements $y_{ij}$ about state deviations to neighboring agents $j\in \Jc_i$. In summary, the agents thus hold $2m=12$ measurements. However, due to the symmetries $y_{ij}=-y_{ji}$, a centralized analysis only needs to consider one measurement per edge. These six measurements form the vector $\yb$, and we define $\yb = (y_{12}\,\, y_{13}\,\, y_{14}\,\, y_{23}\,\, y_{35}\,\, y_{45})^\top$ based on a lexicographical order. The corresponding incidence matrix is given by
	$$
	\Bb = \left(\begin{array}{rrrrrr}
		1 &  1 &  1 &  0 &  0 &  0 \\
		-1 &  0 &  0 &  1 &  0 &  0 \\
		0 & -1 &  0 & -1 &  1 &  0 \\
		0 &  0 & -1 &  0 &  0 &  1 \\
		0 &  0 &  0 &  0 & -1 & -1
	\end{array}\right).
	$$
	Remarkably, the incidence matrix defines a directed subgraph depicted in Figure~\ref{subfig:ex-graph-directed}. Now, in order to specify the paths required for resets, we define another subgraph such that each follower node $i \in \{2,\dots,5\}$ has exactly one incoming edge. A suitable choice, which can be described by
	$$
	\Pb^\top = \left(\begin{array}{c}
		\zerob_6^\top \\
		\pb_2^\top \\
		\vdots \\
		\pb_5^\top
	\end{array}\right) = \left(\begin{array}{cccccc}
		0 & 0 & 0 & 0 & 0 & 0 \\
		1 & 0 & 0 & 0 & 0 & 0 \\
		0 & 1 & 0 & 0 & 0 & 0 \\
		0 & 0 & 1 & 0 & 0 & 0 \\
		0 & 1 & 0 & 0 & 1 & 0 \\
	\end{array}\right),
	$$
	as in~\eqref{eq:pathMatrix}, is shown in Figure~\ref{subfig:ex-graph-directed-sub}. Finally, this graph can also be interpreted as the tree of height $h=2$ (depicted in Figure~\ref{subfig:ex-graph-directed-sub-tree}), which is traversed bottom up during data collection and top down during resets. Studying computation results for this example is not very insightful. Hence, we investigate  more challenging setups next.
	
	\subsection{Medium-scale simulations}
	\label{subsec:sim-medium-scale}
	To investigate the performance of the proposed method, we randomly  generated 1,000 graphs with different dimensions. More precisely, $n$ was picked at random between $10$ and $100$ alongside an ``edge probability'' $p_\mathrm{edge}\in [0.1, 0.7]$. A random graph was then generated by inserting an edge $\{i,j\}$ with probability $p_\mathrm{edge}$ for all $i < j$. To each edge, we assigned a standard deviation $\sigma_{ij}$ from the set $\{0.1, 0.5, 0.9\}$ with equal probability and sampled a corresponding noise realization $v_{ij}$ from $\Nc(0,\sigma_{ij}^2)$. Next, we randomly generated original states $x_i$ in the interval $[-10,10]$ for each state and computed the measurements~\eqref{eq:measurements} for each edge. To build up the required subgraphs, we selected a feasible matrix $\Bb$ and identified the tree using the TAG algorithm from~\cite{madden2002tag}. Regarding $\alpha$, we chose 
	$$
	\alpha = \frac{2}{\lambda_1(\Lb)+\lambda_{n-1}(\Lb)}
	$$
	as suggested in \cite[Eq.~(16)]{xiao2004fast}. For the simulation, we then picked $\kIter \in \{5,10,15\}$ with equal probability and simulated six computation rounds with five intermediate resets for both $w=0$ (soft reset) and $w=n-1$ (hard reset). The encrypted arithmetic was carried out with $s=10^3$ and $q=2^{2048}$ in all 1,000 cases. 
	
\begin{figure*}[ht!]
		\centering
		\hspace*{-5mm}
		\subcaptionbox{\label{subfig:ex-large-soft}}[0.45\textwidth]{
			\includegraphics[trim={{35mm} {107mm} {40mm} {107mm}}, clip, width=\columnwidth]{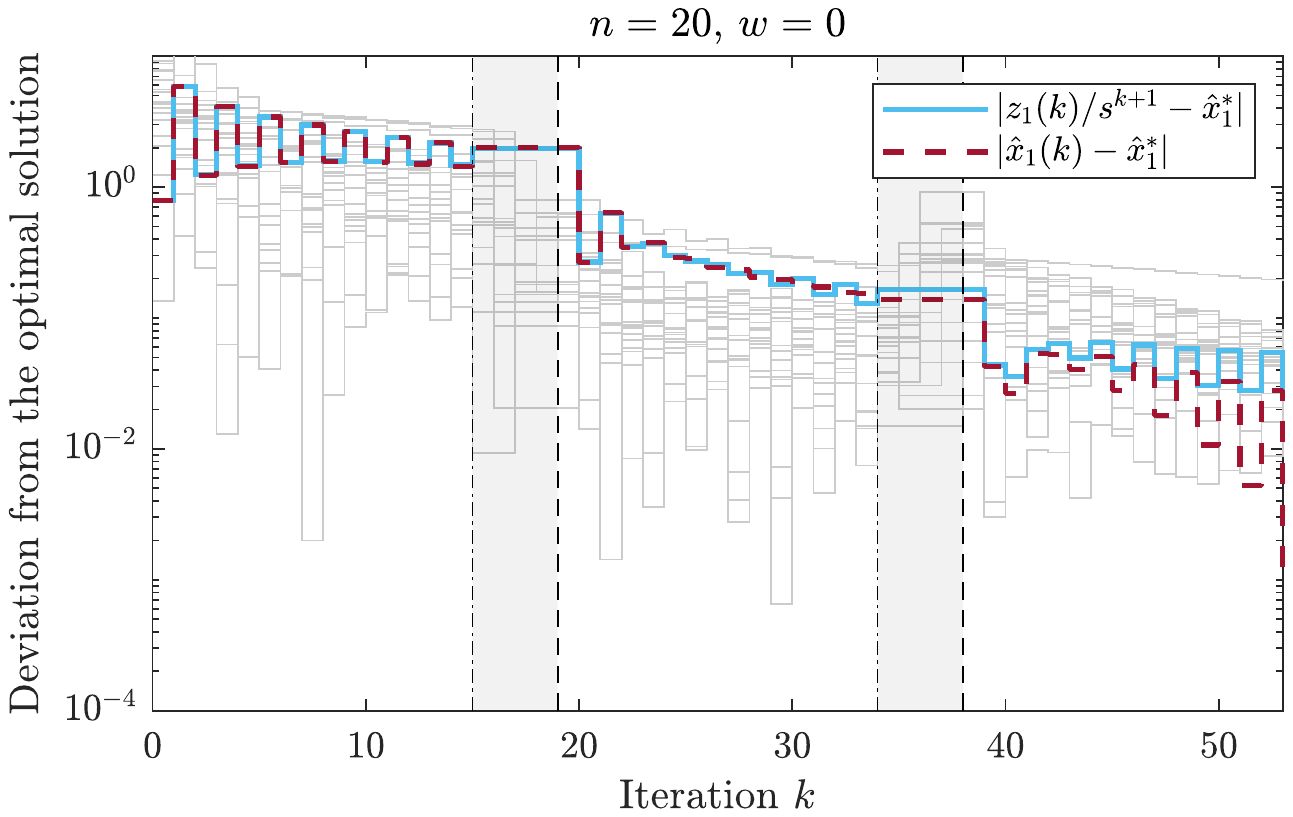}
		}
		\hspace{8mm}
		\subcaptionbox{\label{subfig:ex-large-hard}}[0.45\textwidth]{
			\includegraphics[trim={{35mm} {107mm} {40mm} {107mm}}, clip, width=\columnwidth]{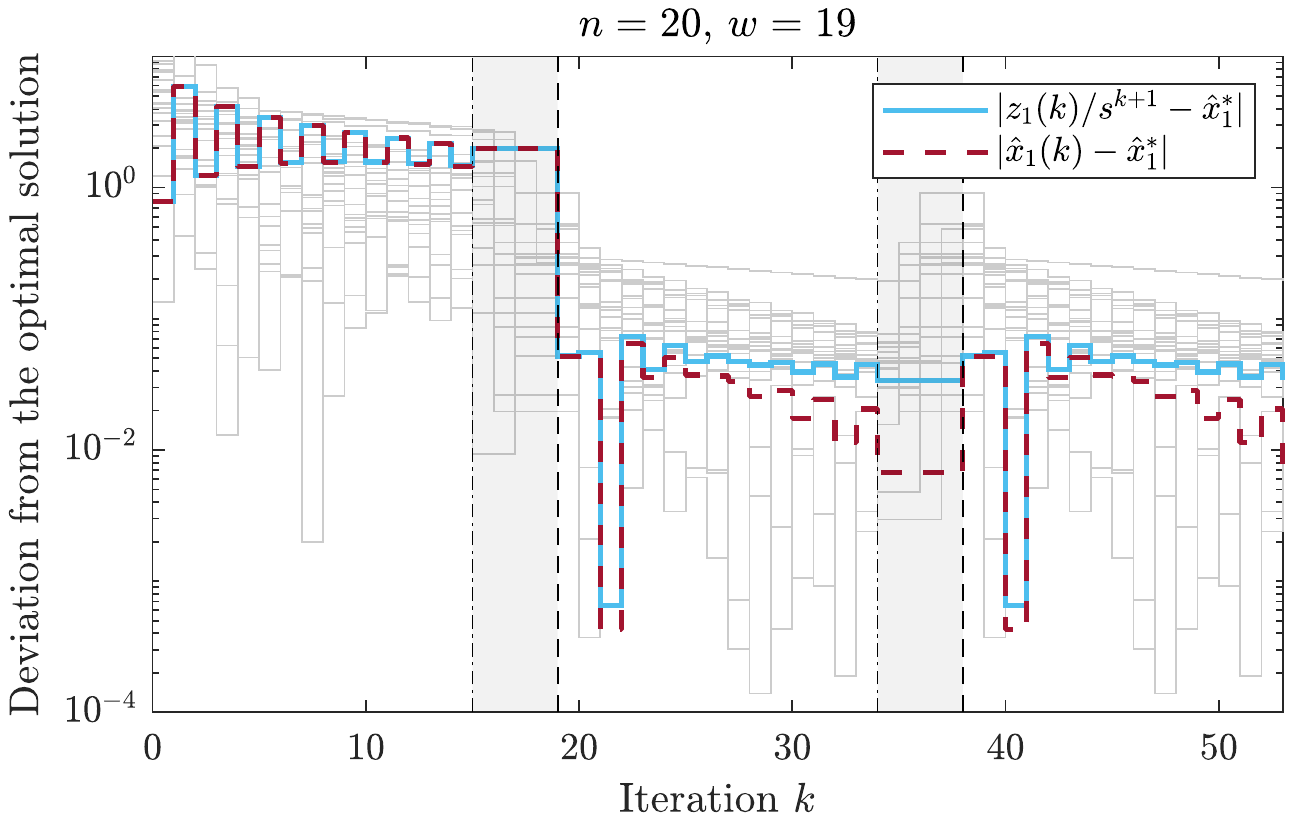}
		}
		\caption{Deviation from the optimal solution of the integer-based and the floating-point implementation for a randomly generated example. For presentation purposes, only the first three iteration phases are shown. The dashed and dash-dotted vertical lines mark the beginnings and ends of reset phases, respectively. (\subref{subfig:ex-large-soft}) shows the soft reset ($w=0$) while (\subref{subfig:ex-large-hard}) shows the hard reset ($w=n-1$). The follower trajectories are shown in gray, and the leader trajectories are shown in color.}
		\label{fig:example-large}
	\end{figure*}

	We consistently observed  behavior similar to that in Figure~\ref{fig:example-large}. Most importantly, the estimates converge towards the optimal solution and the integer-based implementations modulo $q$ and floating-point implementations closely match. In fact, we never observed an integer overflow in $\Z_q$ (for the leader) and the sufficient condition~\eqref{eq:kIterCondition} was always satisfied for $\bar{x}_1=10^4$. Surprisingly, the hard reset, in comparison to the soft resets, often resulted in a faster reduction of the estimation error. However, at the end of the simulations, the hard and soft resets both yield deviations in the order of the achievable accuracy (less than $10^{-2}$) from the optimal solution in about $50\,\%$ of the cases. This happens slightly more often for the soft reset.
	
	\section{Conclusion and outlook}
	\label{sec:outlook}
	In this paper, a novel homomorphically encrypted affine averaging algorithm is presented. The counterweight to many advantages that come with an efficient homomorphic implementation is that only a limited amount of operations can be supported.
	This conflicts with the iterative nature of affine averaging and makes it hard to ensure a sufficiently converged solution a priori. 
	
	We resolve this problem by periodic resets with the help of a leader agent, who has access to its plaintext state in combination with a communication subgraph. This enables us to reset the other agents' states by means of the encrypted leader state and measurements, respectively, which are propagated through the network. As a result, an unlimited runtime of the algorithm is enabled and privacy of the follower agents (which are not neighbors of the leader) is provided. Next, we ensure overflow-free encrypted operation and equivalence with the plaintext analogue by overestimating quantization errors. In order to find an optimal balance between the use of the leader's state and the measurements in the reset strategy, a parameterization is introduced. Here, two cases stand out, where the leader state remains untouched and where it is completely replaced by measurement information. Surprisingly, the different types of resets often result in quite similar behavior (see, e.g., Fig.~\ref{fig:example-large}).
	
	For future research, we will study the implications of the different resets more rigorously. To this end, we will carefully investigate the corresponding covariance matrices in relation to the noise-optimal solution. Finally, time-varying measurements or communication graphs could be a challenging extension.
	
	\section*{Acknowledgement}
	The majority of this work was carried out during an academic visit of Junsoo Kim at TU Dortmund University. Financial support by the German Research Foundation (DFG) and the Daimler and Benz Foundation under the grants SCHU 2940/4-1 and 32-08/19 is gratefully acknowledged.

\end{document}